\documentclass[journal=jacsat,manuscript=article, layout=twocolumn]{achemso}

\usepackage{graphicx}
\usepackage{amssymb}
\usepackage{lineno}
\usepackage{color}
\usepackage{xspace}
\usepackage{amsmath}
\usepackage{hyperref}
\usepackage{tabularx}
\usepackage{chemformula}
\usepackage[T1]{fontenc}

\DeclareRobustCommand{\rchi}{{\mathpalette\irchi\relax}}
\newcommand{\irchi}[2]{\raisebox{\depth}{$#1\chi$}}

\newcommand{\NZMO}{NiZnMo$_{3}$O$_{8}$\xspace}
\newcommand{\FMO}{Fe$_{2}$Mo$_{3}$O$_{8}$\xspace}
\newcommand{\TC}{TeCl$_{4}$\xspace}
\newcommand{\fourMeMO}{M$_{2}$Mo$_{3}$O$_{8}$\xspace (M~=~Mn, Fe, Co, Ni)\xspace}
\newcommand{\fourMeMOshort}{M$_{2}$Mo$_{3}$O$_{8}$\xspace}

\newcommand{\FZMO}{FeZnMo$_{3}$O$_{8}$\xspace}

\newcommand{\NMO}{Ni$_{2}$Mo$_{3}$O$_{8}$\xspace}

\author{Hiraka Haruhiro}
\affiliation{Center for Integrated Nanostructure Physics, Institute for Basic Science (IBS), Suwon 16419, Republic of Korea}
\alsoaffiliation{Sungkyunkwan University, Suwon 16419, Republic of Korea}

\author{Raktim Datta}
\affiliation{Center for Van der Waals Quantum Solids, Institute for Basic Science (IBS), Pohang 37673, Republic of Korea}

\author{Poonam Yadav}
\affiliation{Center for Integrated Nanostructure Physics, Institute for Basic Science (IBS), Suwon 16419, Republic of Korea}
\alsoaffiliation{Sungkyunkwan University, Suwon 16419, Republic of Korea}
\alsoaffiliation{Center for Van der Waals Quantum Solids, Institute for Basic Science (IBS), Pohang 37673, Republic of Korea}

\author{Anzar Ali}
\affiliation{Center for Integrated Nanostructure Physics, Institute for Basic Science (IBS), Suwon 16419, Republic of Korea}
\alsoaffiliation{Sungkyunkwan University, Suwon 16419, Republic of Korea}

\author{Suheon Lee}
\affiliation{Center for Integrated Nanostructure Physics, Institute for Basic Science (IBS), Suwon 16419, Republic of Korea}
\alsoaffiliation{Sungkyunkwan University, Suwon 16419, Republic of Korea}

\author{Matthias J. Gutmann}
\affiliation{ISIS Facility, Rutherford Appleton Laboratory, Chilton, Didcot, OX11 0QX, UK}

\author{Duhee Yoon}
\affiliation{Center for 2D Quantum Heterostructures, Institute for Basic Science (IBS), Suwon 16419, Republic of Korea}
\alsoaffiliation{Sungkyunkwan University, Suwon 16419, Republic of Korea}

\author{Dirk Wulferding}
\affiliation{Department of Physics and Astronomy, Sejong University, Seoul 05006, Republic of Korea}

\author{Xianghan Xu}
\affiliation{Keck Center for Quantum Magnetism, Rutgers University, Piscataway, New Jersey 08854, USA}
\alsoaffiliation{Department of Physics and Astronomy, Rutgers University, Piscataway, New Jersey 08854, USA}

\author{Moon-Ho Jo}
\affiliation{Center for Van der Waals Quantum Solids, Institute for Basic Science (IBS), Pohang 37673, Republic of Korea}

\author{Sang-Wook Cheong}
\affiliation{Keck Center for Quantum Magnetism, Rutgers University, Piscataway, New Jersey 08854, USA}
\alsoaffiliation{Department of Physics and Astronomy, Rutgers University, Piscataway, New Jersey 08854, USA}

\author{Sungkyun Choi}
\affiliation{Center for Integrated Nanostructure Physics, Institute for Basic Science (IBS), Suwon 16419, Republic of Korea}
\alsoaffiliation{Sungkyunkwan University, Suwon 16419, Republic of Korea}
\alsoaffiliation{Center for Van der Waals Quantum Solids, Institute for Basic Science (IBS), Pohang 37673, Republic of Korea}
\email{sungkyunchoi@ibs.re.kr}

\title{Controlled growth of polar altermagnets via chemical vapor transport}

\abbreviations{ME}

\begin{document}

\begin{abstract}
Altermagnetic properties have been recently proposed in polar magnetic oxides, \fourMeMO, where improved characteristics of stronger magnetoelectric coupling and higher magnetic transition temperatures were observed. Thus, understanding their microscopic origins is of fundamental and technological importance. However, the difficulty in growing large single crystals hinders detailed experimental studies. Here, we report the successful growth of large single crystals of the pyroelectric antiferromagnet using two representative compounds, \FMO and \NZMO. Growth was optimized using various parameters, finding the transport agent density as a primary factor, which depends strongly on the position of the pellet, the starting powder form, and the volume of the ampule. We demonstrated a controlled growth method by manipulating the convection and diffusion kinetics. High-quality crystals were characterized by using single-crystal X-ray diffraction, Laue diffraction, magnetic susceptibility, and Raman spectroscopy. Manipulation of magnetic properties through nonmagnetic Zn doping was shown in \NZMO. Our results enable the detailed investigation and manipulation of their unconventional altermagnetic and multiferroic properties. This study provides crucial insight into the controlled growth of other functional quantum materials.
\end{abstract}

\section{Introduction}
\label{intro}
Multiferroic (ME) materials~\cite{Cheong2007, Spaldin2019} are traditionally categorized into the so-called type-I and type-II~\cite{Khomskii2009}. Type-I multiferroics comprise materials that exhibit independent ferroelectricity and magnetism, as observed in BiFeO$_{3}$~\cite{Wang2003} and YMnO$_{3}$~\cite{Aken2004}. In these systems, ferroelectricity occurs at temperatures higher than the magnetic transition temperature, indicating weaker coupling between the electric polarization and the magnetic moment. On the other hand, the ferroelectricity is induced by magnetism in the type-II multiferroics. This implies a strong coupling between the two, which has been studied in TbMnO$_{3}$~\cite{Kimura2003} and TbMn$_{2}$O$_{5}$~\cite{Hur2004}. However, their magnetism appears at very low temperatures, accompanied by a much weaker electric polarization. Thus, the practical application of both classes of materials is fundamentally challenging.

In this regard, honeycomb-based pyroelectric compounds \fourMeMO~\cite{McCarroll1957} have garnered significant attention due to their unconventional magnetoelectric (ME) properties, which offer improved magnetoelectric performance compared to the much-studied multiferroic materials. They are considered a novel class of multiferroics, referred to as type-III~\cite{Shankar2020}. These compounds possess the advantages of stronger magnetoelectric coupling than type-I and higher magnetic transition temperatures than type-II multiferroics. Moreover, polar magnets favor single magnetic domains, eliminating the need for poling procedures~\cite{Wang2015} due to their noncentrosymmetric crystal structures.

Numerous intriguing ME phenomena are reported in this compound family. Examples include a large phonon magnetic moment enhanced by magnetic fluctuations in \FMO~\cite{Wu2023}, hybridized band-Mott gaps in \FMO~\cite{Park2021, Stanislavchuk2020}, tunable magnetoelectricity~\cite{Kurumaji2015}, giant thermal Hall effects~\cite{Ideue2017}, optical diode effects~\cite{Yu2018}, axion-type coupling~\cite{Kurumaji2017:FZMO} in \FZMO, and nonlinear ME properties associated with a noncollinear magnetic order in \NMO~\cite{Tang2021, Yadav2023}.

Very recently, \fourMeMO are also recognized by altermagnets~\cite{Mazin2022} via magnetic point group analysis~\cite{Cheong2024:arxiv}. Altermagnetism is defined as the state where spins are fully compensated, but the parity-time reversal symmetry is broken~\cite{Mazin2022}, which is further extended to noncollinear antiferromagnets~\cite{Cheong2024}. This new magnetism has attracted tremendous interest with the prospect of observing various intriguing spin-related phenomena. These include anomalous Hall effects~\cite{Cheong2024:order, Nagaosa2009}, piezomagnetism~\cite{Ikhlas2022, Radaelli2024}, and current-induced magnetization~\cite{Cheong2024:Kinemagnet}, which originate from the spin-split band structure in momentum space~\cite{Cheong2024:arxiv}. Thus, discovering and exploring those phenomena in altermagnets are contemporary research topics. Also, understanding the relationship between type-III multiferroics and altermagnetism will be of significant importance. Thus, detailed experimental investigations and comparisons with theoretical predictions are required. To pursue these goals, systematically growing large single crystals of \fourMeMOshort is essential.

Herein, we report the synthesis of large single crystals of two pyroelectric honeycomb magnets \FMO and \NZMO, via controlled chemical vapor transport. We optimized growth parameters over a wide range, enabling the synthesis of large crystals. The transport agent density was identified as a primary growth parameter, strongly influenced by the position of the pellet, the starting powder form, and the volume of the ampule. Optimized growth conditions for each compound were determined using two different experimental setups. Comprehensive characterization, including manipulated magnetic properties, confirms the high quality of the synthesized crystals. This work provides helpful information that can be utilized in growing altermagnetic single crystals and other functional quantum materials.

\begin{figure*}[t]
\begin{center}
\includegraphics[width=\linewidth]{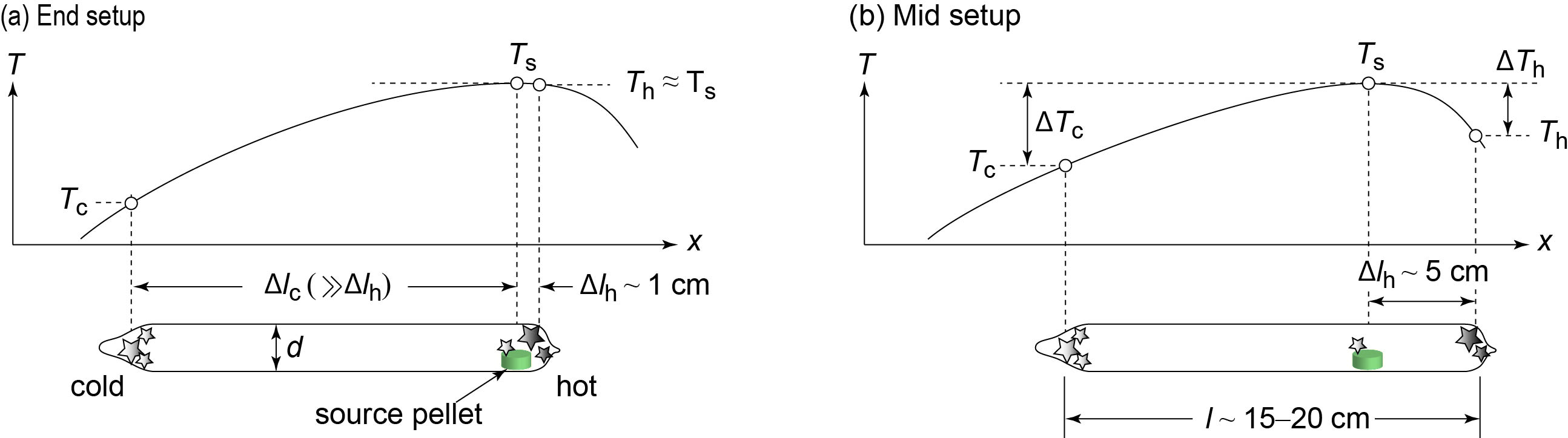}
\end{center}
\caption{Two growth configurations used in this study: (a) end setup and (b) mid setup. They depend on the relative position of the pellet within the quartz tube along the horizontal direction. A curved line denotes the actual temperature profile inside the furnace.
}
\label{setup}
\end{figure*}

\section{Experimental Details}
\label{experimetal}

\subsection{Growth Procedure}
\label{procedure}
The growth of single crystals is fundamental to advancing various fields, including condensed matter physics, chemistry, and materials science engineering. Among many tools, the chemical vapor transport (CVT) method~\cite{May2020} is a versatile growth technique and has thus been widely used to synthesize quantum materials, such as van der Waals materials~\cite{Wang2022} and ME~\cite{Cheong2007, Spaldin2019} systems. The sample growth can generally be understood by thermodynamics and kinetics, with the latter being relatively less examined in the literature.

In this study, we utilized the CVT growth method in a kinetically controlled way. To prepare for the growth, starting materials, including Fe$_2$O$_3$ (99.9\%, Kojundo Chemical Laboratory Co., Ltd.), NiO (99\%, Alfa Aesar Fine Chemicals \& Metals), ZnO (99.99\%, Kojundo Chemical Laboratory Co., Ltd.), Mo (99.9\%, Alfa Aesar), and MoO$_3$ (99.5\%, Alfa Aesar Fine Chemicals \& Metals), were mixed thoroughly in stoichiometric ratios and pelletized. Raw powder (labeled 0S) of \FMO and \NZMO and polycrystalline (1S) \NZMO samples were used. The 1S sample was obtained by sintering the 0S pellet in a vacuum at 950~$^{\circ}$C for 6 h, followed by grinding and repelletizing. Each pellet had a diameter of 1~cm (mass of 1.5--2.0~g). \TC was used as the transport agent~\cite{Strobel1982, Strobel1983} in amounts ranging from 0.02 to 0.17~g. The source pellet was sealed with \TC powder in a quartz ampule under vacuum conditions. The transport agent density is defined as $\rho_{\rm a} = m_{\rm a} / V$, where $m_{\rm a}$ is the transport agent mass and $V$ is the ampule volume. Two ampule types, with inner diameters of ($d$) $1.2$ and $1.6$~cm, were used. The ampules were 2~mm thick with lengths ($l$) of 15--20~cm (Figure~\ref{setup}). The sealed quartz ampules were placed in a horizontal two-zone furnace for the crystal growth. The left zone was maintained at 900$~^{\circ}$C and the right zone at 1,000$~^{\circ}$C. The furnace's temperature was fixed to better understand the kinetics of crystal growth. The sealed ampules were heated to the target temperature for 24 h and maintained for 2 weeks before cooling. N$_2$ gas ($\sim$10 cc/min) was used during the growth process to minimize possible oxidation through microcracks in the quartz ampules. We synthesized the crystals in over 100 batches to obtain reliable conclusions.

\subsection{Two Growth Configurations}
\label{setups}
Figure~\ref{setup} illustrates two growth configurations used in this study. We controlled the convection and diffusion growth at the cold and hot positions of the ampule using two configurations: ``end'' and ``mid'' setups (Figure~\ref{setup}). $T_{\rm c}$ ($T_{\rm_h}$) is the temperature at the ampule's cold (hot) position at the two ends. $T_{\rm s}$ is the temperature at the source position, which corresponds to the location of the maximum temperature in the furnace.

Figure~\ref{setup}a shows the end setup, wherein the source pellet was very close to the hot position of the ampule but far from the cold position. The temperature difference between the pellet and the cold position was thus large, while the pellet and hot zone temperatures were nearly the same. Because the pellet is very close to the hot position, diffusion growth is expected to dominate at the hot position, whereas convection growth dominates at the cold position.

Figure~\ref{setup}b illustrates the mid setup, wherein the source pellet was moved approximately 5 cm from its previous position. As a result, the distance between the source and the cold position ($\Delta l_{\rm c}$) decreased, whereas the distance between the source and the hot position ($\Delta l_{\rm h}$) increased. Thus, convection growth is expected to be suppressed in the mid setup compared to the end setup. After the growth, the relative effects of convection and diffusion kinetics on the sample growth were evaluated by comparing the mass of the largest single crystal and the total mass of the grown crystals in each batch.

\begin{figure*}[t]
\begin{center}
\includegraphics[width=\linewidth]{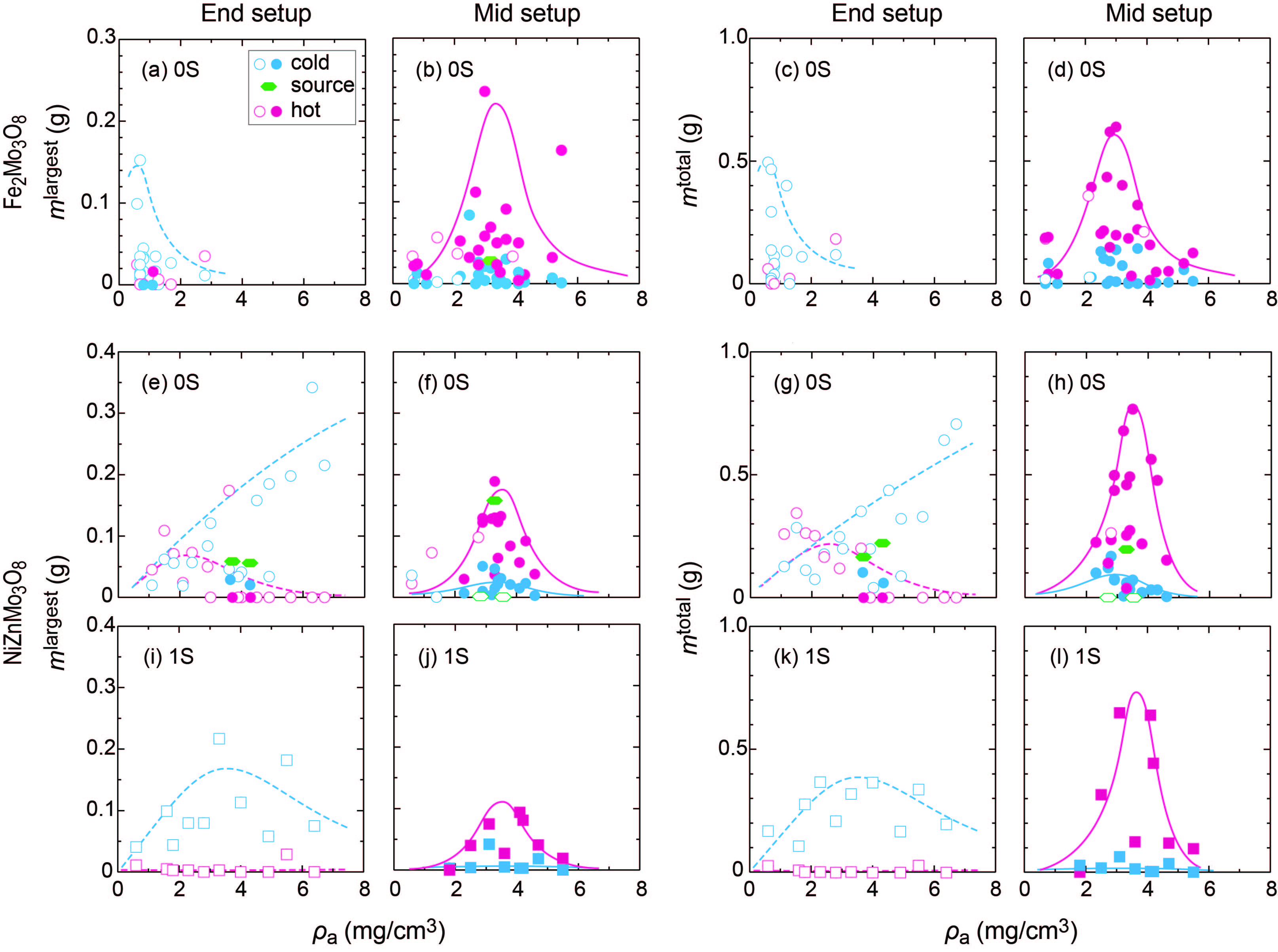}
\end{center}
\caption{Single-crystal growths results: (a-d) \FMO and (e-l) \NZMO. $m^{\rm largest}$ and $m^{\rm total}$ are the masses of the largest single crystal and the total masses of the grown crystals per ampule, respectively. The blue, green, and red symbols denote the growth results from the cold, hot, and source positions, respectively. Open (closed) symbols denote a larger (smaller) diameter of the ampule with the corresponding dashed (solid) lines for guidance, i.e., $d = 1.6$ ($1.2$)~cm.
}
\label{fmo_nzmo}
\end{figure*}

\begin{figure} [t]
\begin{center}
\includegraphics[width=\linewidth]{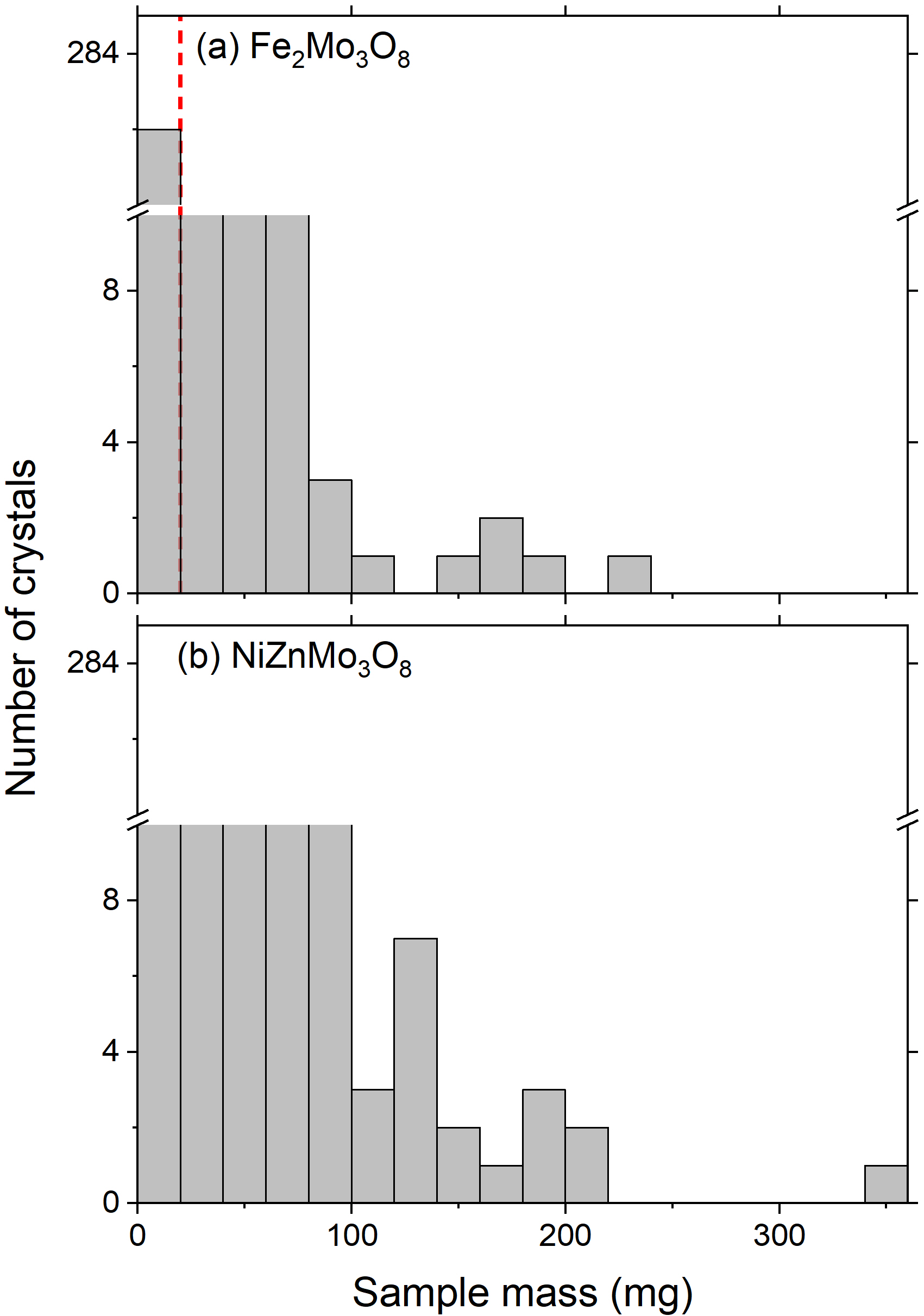}
\end{center}
\caption{Histograms of the number of grown crystals: (a) \FMO and (b) \NZMO. A vertical dashed red line in (a) is the typical size of the single crystals used in previous studies.
}
\label{hist}
\end{figure}

\subsection{Characterization}
\label{char}
The qualities and properties of the grown \FMO and \NZMO single crystals were characterized using various techniques. Shiny, clean, tiny crystals were selected for single-crystal X-ray diffraction (XRD) measurements. They were mounted on a microloop with grease. Diffraction data were collected at room temperature using the XtaLAB Synergy S with Mo--K$_{\alpha}$ radiation.

X-ray Laue diffraction was also performed using an X-ray generator operating at 30 kV and 20 mA. Two crystals, \FMO ($\sim$44~mg, $\sim$2 $\times$ 2 $\times$ 1.2 mm$^{3}$) and \NZMO ($\sim$34~mg, $\sim$2.2 $\times$ 2 $\times$ 1.0 mm$^{3}$), were used for the measurements. The data were collected over a few minutes.

Zero-field-cooled (ZFC) and field-cooled (FC) magnetic susceptibility measurements were performed by using a Physical Property Measurement System (Dynacool, Quantum Design) equipped with a vibrating sample magnetometer module.

Raman spectra were acquired using a 532 nm wavelength solid-state laser (Cobolt Samba-500 series). The laser was focused on the samples with a spot diameter of 50 $\mu$m and an acquisition time of 20 s, using a Horiba iHR-320 spectrometer equipped with a Synapse charge-coupled device (CCD) detector. For cross-checks, additional Raman measurements were performed using a 514.5 nm Ar-ion laser as the excitation source. The laser beam was focused on the samples through a 20$\times$ objective lens in backscattering geometry. Raman signals were recorded with a liquid-nitrogen-cooled CCD detector using a Horiba iHR550 spectrometer. All measurements were performed at room temperature.

\begin{figure} [t]
\begin{center}
\includegraphics[width=\linewidth]{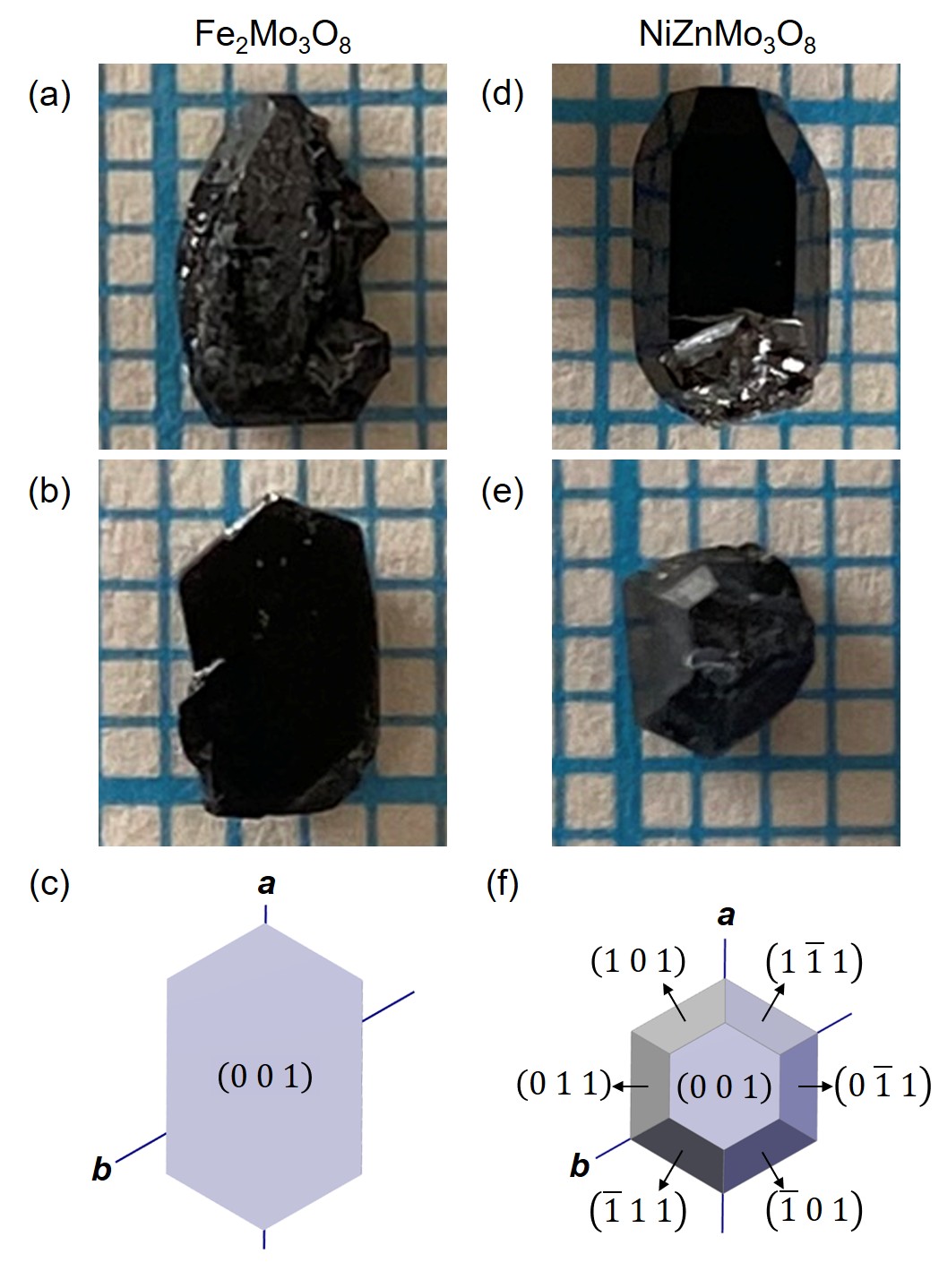}
\end{center}
\caption{Representative large single crystals for (a--c) \FMO and (d--f) \NZMO. The grid length is 1~mm. Two typical crystal morphologies illustrated in parts c and f are based on (b) and (e). Their masses are 65 and 62~mg for (a) and (b), and 157 and 84~mg for (d) and (e), respectively.
}
\label{xtals}
\end{figure}

\section{Results and Discussion}
\label{results}
Figure~\ref{fmo_nzmo} shows the growth results. Results for \FMO using the raw powder pellet (labeled 0S) are presented in Figures~\ref{fmo_nzmo}a--d. In the end setup, single crystals predominantly formed at the cold position [blue empty circles in Figure~\ref{fmo_nzmo}a]. Large crystals were synthesized at $\rho_{\rm a}^{\rm end} \lesssim 1$~mg/cm$^{3}$ [blue empty circles in Figure~\ref{fmo_nzmo}a], which is close to the previously reported $\rho_{\rm a}=1.45$~mg/cm$^{3}$~\cite{Strobel1982}. On the other hand, larger single crystals were grown at the hot position in the mid setup; i.e., an optimal value of $\rho_{\rm a}$ was $\sim$3~mg/cm$^{3}$ [red solid circles in Figure~\ref{fmo_nzmo}b]. The large single crystals were synthesized by using larger ampules (1.6 cm in diameter) in the end setup, indicating that larger ampules accelerated convection-driven growth at the cold position. Conversely, smaller ampules (1.2 cm in diameter) favored growth at the hot position in the mid setup due to their shorter transport distance, resulting in the formation of larger single crystals. Figure~\ref{fmo_nzmo}c,d presents the total mass of grown crystals, revealing a similar trend.

Figures~\ref{fmo_nzmo}e--h shows the \NZMO growth using the raw powder (0S). Figure~\ref{fmo_nzmo}e reveals an optimized $\rho_{\rm a}$ value likely just beyond the measured range at approximately 9~mg/cm$^{3}$. Medium-sized crystals ($0.05$~g $\lesssim m^{\rm largest} \lesssim 0.1$~g) were formed at the hot position for lower $\rho_{\rm a}$ values [red empty symbols in Figure~\ref{fmo_nzmo}e]. In the end setup, one large single crystal ($m^{\rm largest}$ $\le$ $\sim$0.35~g) was usually formed per ampule [blue empty circles in Figure~\ref{fmo_nzmo}e]. In the mid setup, single crystals were preferably grown at the hot location [red solid circles in Figure~\ref{fmo_nzmo}f]. The optimized $\rho_{\rm a}$ value is approximately $3$~mg/cm$^{3}$ [a solid red line in Figure~\ref{fmo_nzmo}f], which is significantly lower compared to that of the end setup [a dashed blue line in Figure~\ref{fmo_nzmo}e]. The end setup favors the growth of one large crystal, while the mid setup is more suited for growing multiple medium-sized crystals of \NZMO.

Further, we used the polycrystalline \NZMO powder sample (1S) to examine its effect on single crystal growth. The results were similar to those of the raw powder case (0S). However, there is a main difference is the decreased optimal $\rho_{\rm a}$ of $\sim$3.5~mg/cm$^{3}$ in the end setup [a dashed blue line in Figure~\ref{fmo_nzmo}i], compared to that of the 0S powder  [a dashed blue line in Figure~\ref{fmo_nzmo}e]. By using the 0S source instead of the polycrystalline 1S source, chemical reactions between reagents and Cl-containing gases needed to grow crystals can increase, where more transport agents can be helpful in producing larger single crystals with 0S. Note that an optimized $\rho_{\rm a}$ in the mid setup is comparable to that for the 0S case [red lines and symbols in Figure~\ref{fmo_nzmo}j,f].

Comparing the results from the two setups (Figure~\ref{fmo_nzmo}e,i,f,j) revealed that the end setup promotes good thermal convection due to a larger temperature difference ($\Delta T_c$) between the pellet and the cold position. We note that single crystals were also grown at the pellet position [green symbols in Figure~\ref{fmo_nzmo}], indicating non-negligible vapor growth~\cite{Stringfellow1991}. The general trend of the total mass of the grown crystals was similar to that of the largest single crystals; i.e., compare the two columns on the left and right panels of Figure~\ref{fmo_nzmo}. We note that growth results using opposite experimental conditions---a smaller (larger) diameter of the ampule in the end (mid) setup---did not reveal a particular trend, as presented in Figure~\ref{fmo_nzmo}. This can be understood by the relatively suppressed convection and diffusion kinetics under the two opposite conditions.

Figure~\ref{hist} displays the distribution of the number of synthesized crystals. It shows that our optimized methods yielded significantly larger single crystals of \FMO compared to the typical crystal sizes used in previous studies~\cite{Wang2015, Stanislavchuk2020, Park2021, Wu2023}; i.e., a typical mass of 20~mg, denoted by a dashed red line in Figure~\ref{hist}a. We successfully grew 10 times larger \FMO crystals in this study; i.e., they were up to $\sim$0.2~g [see Figures~\ref{fmo_nzmo}a,b and Figures~\ref{xtals}a,b].

In this study, we used different quartz ampule volumes (varying diameters), pellet masses, starting powder samples (0S and 1S), experimental setups (end and mid), and compounds (\FMO and \NZMO). Despite the wide range of varying conditions, crystal growth showed a general trend: \TC density establishes the optimized growth [summarized in Figure~\ref{fmo_nzmo}]. This can also be understood by the fact that the source quantity exceeds the transport agent's amount in the CVT growth, and thus the \TC amount effectively determines the transport rate.

Comparing the crystal growth of the two compounds, the growth of large \NZMO single crystals proved more feasible compared to \FMO. This was mainly due to ampule corrosion during \FMO growth, which disrupts the long-term equilibrium conditions required for larger single crystal formation. This difficulty can be avoided by suppressing the convection kinetics while enhancing the diffusion and vapor transport in the mid setup, thereby producing larger \FMO single crystals (Figure~\ref{fmo_nzmo}b). The typical size of the large \NZMO single crystal is even greater (Figure~\ref{fmo_nzmo}e). This was likely because ampule corrosion was not present, allowing the convection growth mechanism to work effectively and maintain longer equilibrium conditions. The end setup with a larger quartz diameter facilitated the growth of sizable \NZMO crystals [empty blue circles in Figure~\ref{fmo_nzmo}e], whereas the mid setup with a smaller quartz diameter facilitated the synthesis of large \FMO crystals [solid red circles in Figure~\ref{fmo_nzmo}b].

Figure~\ref{xtals} presents selected single crystals with well-defined facets in a prism-like shape. Figure~\ref{xtals}a,b, and d,e correspond to \FMO and \NZMO single crystals, respectively. Figure~\ref{xtals}c,f displays typical crystal facets matching those Figure~\ref{xtals}b,e respectively. These revealed the characteristic crystallographic facets that were naturally grown and subsequently indexed.

\section{Characterization}
\label{char}

\begin{figure} [t]
\includegraphics[width=\linewidth]{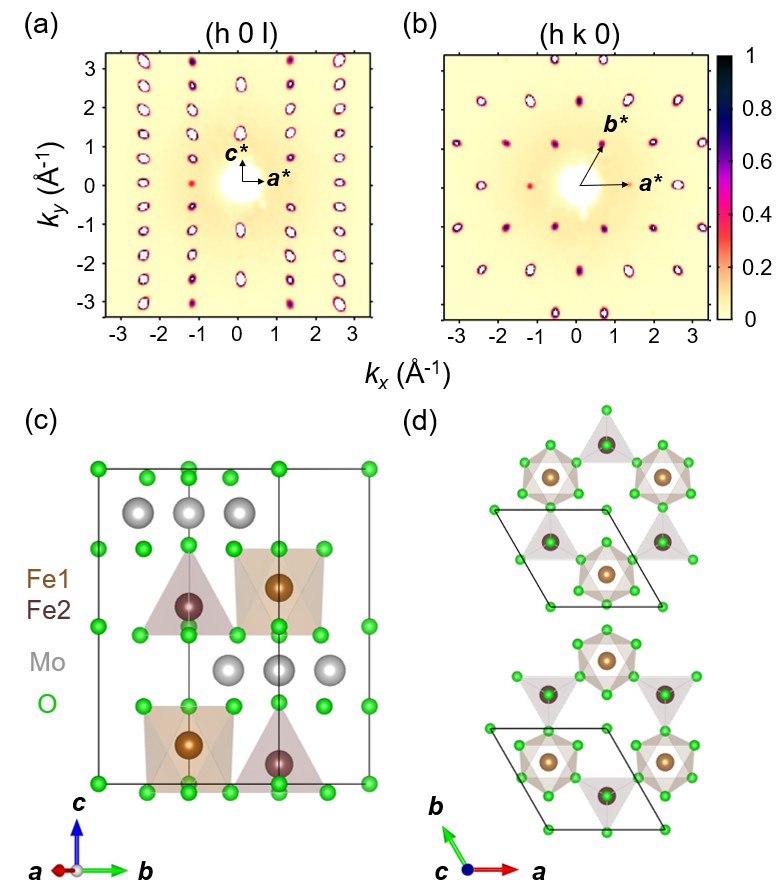}
\caption{XRD intensities for \FMO: (a) (h 0 l) and (b) (h k 0). The crystal structure and the honeycomb layers of \FMO are shown in parts (c) and (d). In (d), the range of the unit cell is 0~$\leq$~$z$~$\leq$~0.3 and 0.4~$\leq$~$z$~$\leq$~0.7 for the top and bottom panels, respectively.
}
\label{sXRD}
\end{figure}

\begin{figure} [t]
\includegraphics[width=\linewidth]{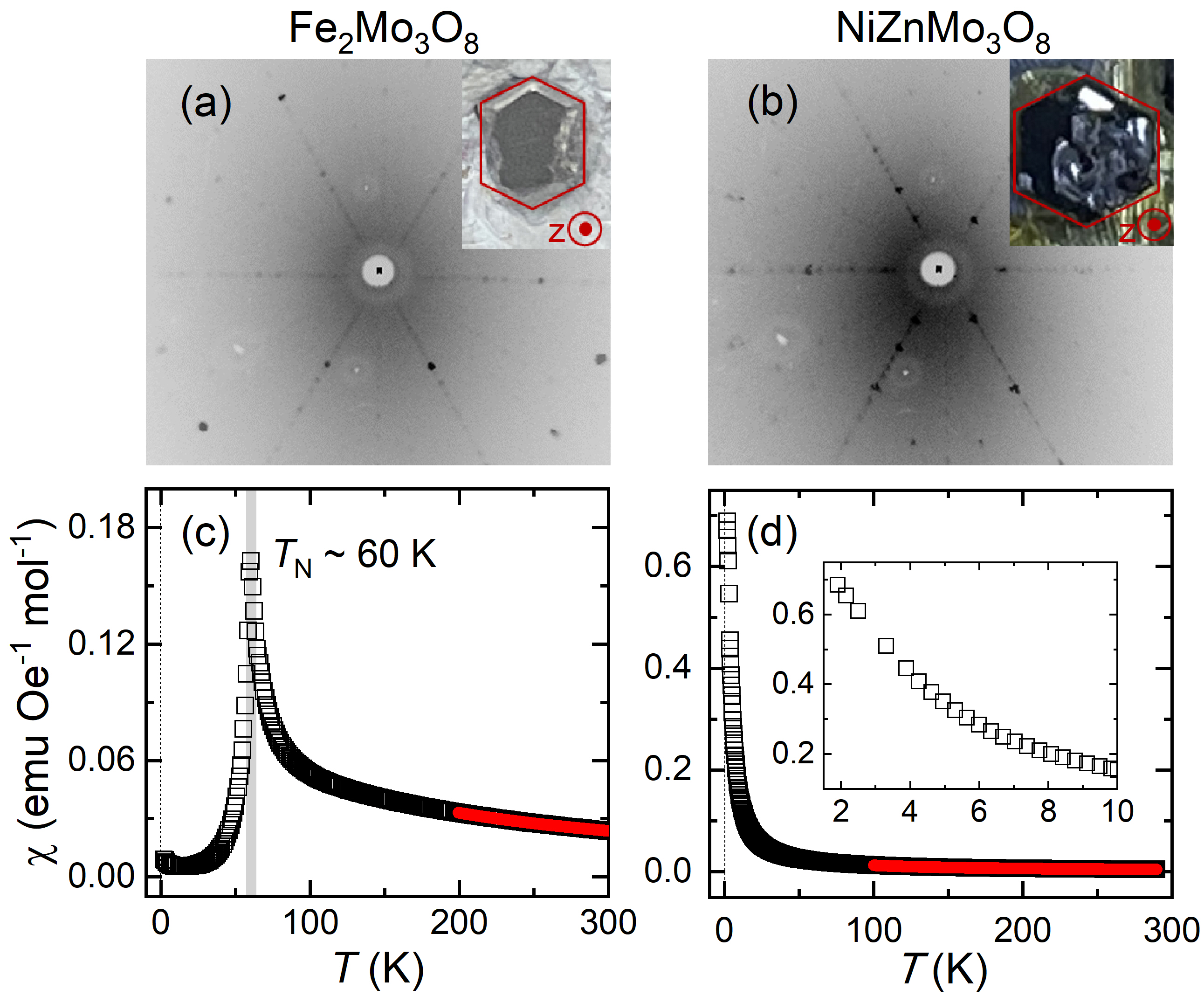}
\caption{(a, b) X-ray Laue diffraction patterns for \FMO and \NZMO single crystals. (c, d) ZFC magnetic susceptibility. The inset of (d) presents a magnified view at lower temperatures. The red lines denote the CW fits (see the texts). The magnetic field of 0.2~T is applied along the $c$-axis.
}
\label{Laue_Chi}
\end{figure}

\subsection{Single-Crystal XRD}
\label{x-ray}
Single-crystal XRD measurements were performed on single crystals of both compounds. We found a single structural hexagonal domain with well-diffracted Bragg peaks. The lattice parameters for \FMO were determined as $a$ = 5.781(1), $c$ = 10.066(1)~\AA~(consistent with a previous report~\cite{Stanislavchuk2020}), whereas for \NZMO, the lattice parameters were $a$ = 5.749(1), $c$ = 9.805(1)~\AA~. Figure~\ref{sXRD}a,b shows two representative reciprocal planes for \FMO, the (h 0 l) and (h k 0) planes. They revealed sharp Bragg peaks in a given crystal symmetry without impurity peaks, indicating the high quality of the crystal. A clear hexagonal peak pattern was also observed in the plane (Figure~\ref{sXRD}(b)). Figure~\ref{sXRD}c,d illustrates the crystal structure of \FMO~\cite{Stanislavchuk2020}. Figure~\ref{sXRD}d displays a rather unique honeycomb lattice composed of FeO$_{6}$ octahedra and FeO$_{4}$ tetrahedra. Two types of Fe ions are chemically distinct because they occupy different Wyckoff sites. This allows the site-specific doping, which we applied in \NZMO, as discussed below.

In \NZMO, Zn is presumed to occupy the tetrahedral sites (Figure~\ref{sXRD}d), as proposed in a similar study of \FZMO~\cite{Streltsov2019}. Zn ions in \FZMO preferentially occupy the tetrahedral sites due to the different ionic radii of Fe and Zn ions. Note that the Zn positioning in \NZMO was not determined via XRD due to the negligible contrast between Ni and Zn X-ray scattering signals. Thus, neutron diffraction studies will be required for determining the Zn position and examining site mixing in \NZMO in the future.

\begin{table} [t]
\begin{center}
\caption{Parameters Obtained from CW fits Using Magnetic Susceptibility $\rchi(T)$. Unit of $\rchi_{0}$, $C$, and $\mu_\textrm{eff}$ are emu~Oe$^{-1}$~mol$^{-1}$, emu~K~Oe$^{-1}$~mol$^{-1}$, and $\mu_\textrm{B}$/Fe ($\mu_\textrm{B}$/Ni), respectively.
}
\vspace{0.4cm}
\label{CW}
\setlength\extrarowheight{10pt}
\setlength{\tabcolsep}{4pt}
\begin{tabular}{c c c c c c c c}
	\hline
 Sample & $\rchi_{0}$($\times$10$^{-4}$) & $\theta$ (K) & $C$ & $\mu_\textrm{eff}$\\
\hline
      \FMO    & 0         & $-$56.18      & 8.465    & 4.12         \\
      \NZMO & 5.50     &     5.05        & 1.160    & 3.04         \\
\hline

\end{tabular}

\end{center}
\end{table}

\subsection{X-ray Laue Diffraction}
\label{laue}
To confirm the orientation and morphology of the grown single crystals, X-ray Laue measurements were performed on \FMO and \NZMO single crystals (Figure~\ref{Laue_Chi}a,b, respectively). These measurements were conducted in the $ab$ plane by using a back-reflection geometry. The Laue patterns of both crystals exhibit hexagonal symmetry, with sharp and well-defined Bragg peaks, confirming the high quality of the crystals. No multidomains were detected. The insets in Figure~\ref{Laue_Chi}a,b shows microscopic images of the \FMO and \NZMO single crystals used, respectively.

\subsection{Magnetic Susceptibility}
\label{magnetic}
Figure~\ref{Laue_Chi}c,d presents the ZFC magnetic susceptibilities of the \FMO and \NZMO single crystals. Note that the FC data were nearly identical. Figure~\ref{Laue_Chi}c reveals a sharp peak at $\sim$60~K, which is a signature of long-range antiferromagnetic order, consistent with the literature~\cite{Wang2015}. The data between 200 and 300~K were fitted using the Curie--Weiss (CW) equation, $\rchi(T) = \rchi_{0} + C/(T-\theta)$, where $\rchi_{0}$ denotes temperature-independent susceptibility (background), $C$ denotes the Curie constant, and $\theta$ denotes the CW temperature. The CW fit yielded $\theta$ = $-$ 56.18~K, indicative of antiferromagnetic interactions. The effective magnetic moment ($\mu_\textrm{eff}$) is $\sim$4.12~$\mu_\textrm{B}$/Fe. This differs from the expected value from the spin-only model (4.9~$\mu_\textrm{B}$/Fe), implying the importance of spin--orbit coupling.

Figure~\ref{Laue_Chi}d shows the temperature-dependent magnetic susceptibility of \NZMO. No clear anomalies were observed down to the lowest temperature of 2~K in these data, necessitating further investigation. The inset of Figure~\ref{Laue_Chi}d shows enlarged data near the base temperature. The suppression of the ordering temperature can be attributed to the weakened exchange interactions in \NZMO compared to \NMO, as nonmagnetic Zn ions likely disrupt nearest-neighbor coupling at the tetrahedral site, as suggested in \FZMO~\cite{Streltsov2019}. A CW fit between 100 and 300~K yielded a positive Curie temperature, suggesting dominant ferromagnetic interactions. The effective moment extracted from this fit is $\sim$3.04~$\mu_\textrm{B}$/Ni. The CW fitting results are summarized in Table~\ref{CW}.

\begin{figure} [t!]
	\includegraphics[width=\linewidth]{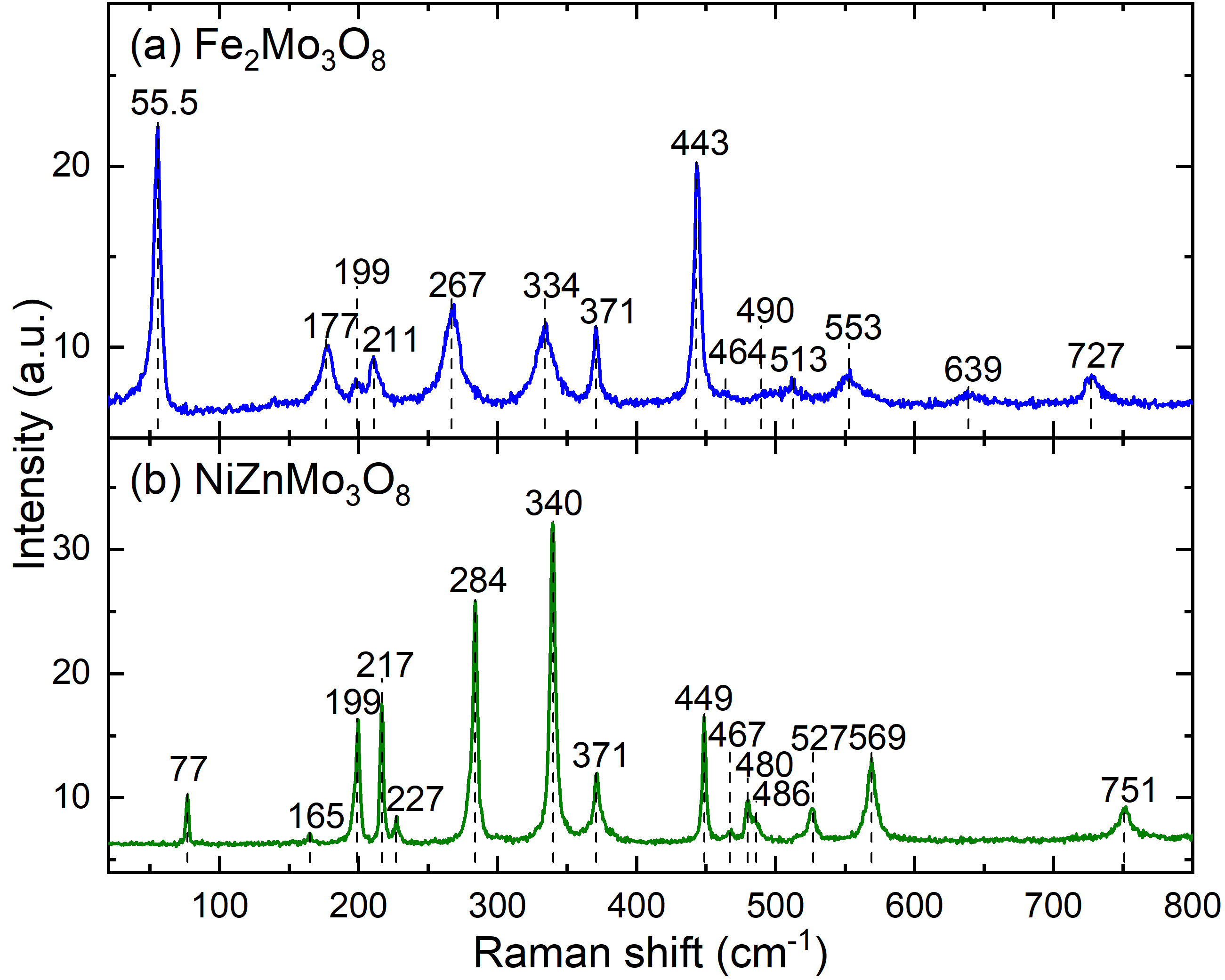}
	\caption{Raman spectra of (a) \FMO and (b) \NZMO. The vertical dashed lines represent the positions of the measured Raman peaks.
	}
	\label{Raman}
\end{figure}

\subsection{Raman Scattering}
\label{raman}
Raman spectra were collected from both \FMO and \NZMO crystals. Figure~\ref{Raman}a,b presents clear and sharp Raman peaks observed for \FMO and \NZMO, respectively [see vertical dashed lines]. This implies the high quality of the grown crystals. Numerous \FMO\ and \NZMO crystals grown in multiple batches exhibited very similar Raman spectra, demonstrating consistently high sample quality across different batches and confirming the reproducibilty of sample growth.

In particular, we observed a prominent phonon at around 55.5 cm$^{-1}$ in \FMO (Figure~\ref{Raman}a). An unusually large magnetic moment was recently proposed on this phonon peak, enhanced in the presence of magnetic fluctuations~\cite{Wu2023}. This can be understood by a hybridized lattice vibration with magnetic order below the N\'{e}el temperature. The observation of a similar Raman peak at 77~cm$^{-1}$ in \NZMO (Figure~\ref{Raman}b) suggests the possibility of a similar coupling mechanism in this compound. Further investigations are needed to explore whether this phenomenon is universal in this compound family.

\begin{figure} [t!]
	\includegraphics[width=\linewidth]{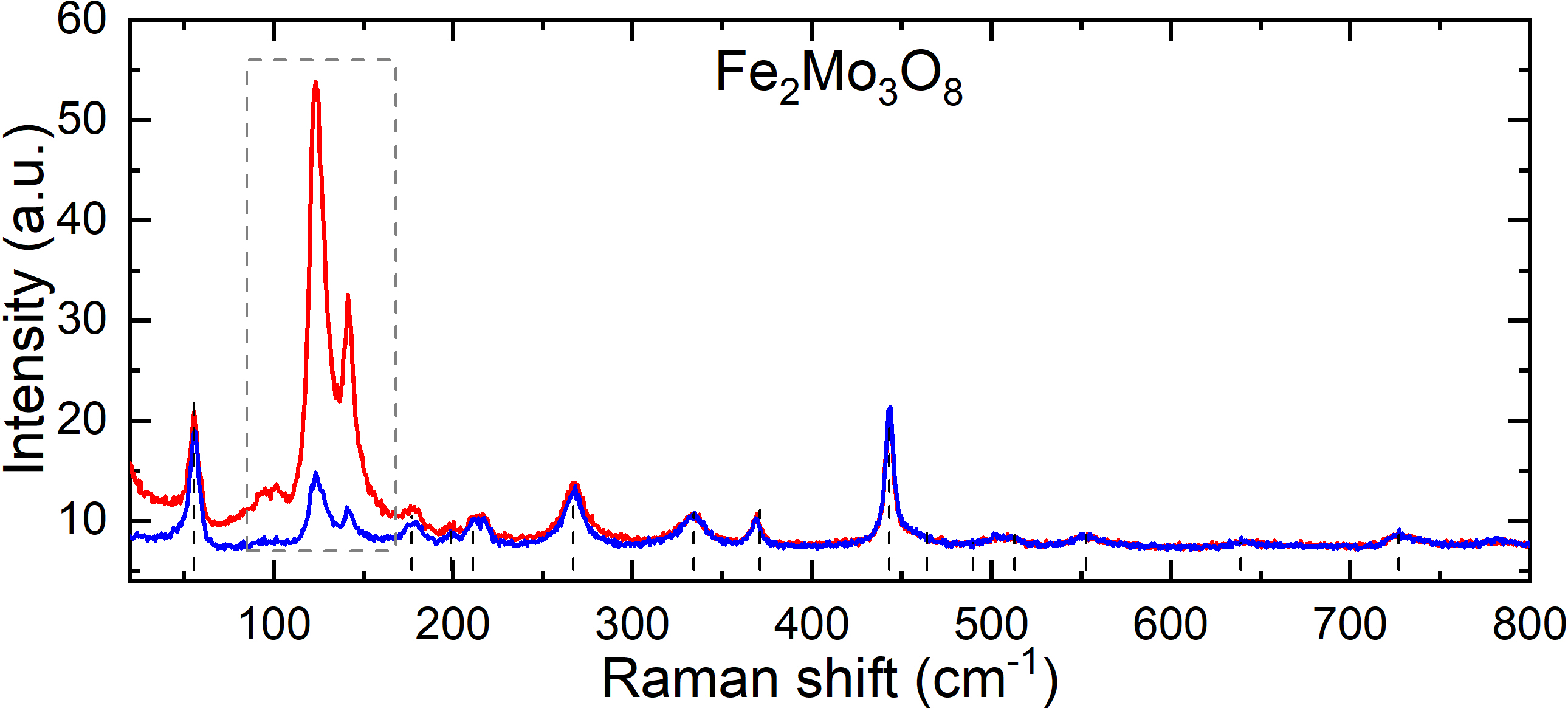}
	\caption{Raman spectra of the \FMO crystal recorded at two different spots. The vertical dashed lines are from Figure~\ref{Raman}a for comparison.
	}
	\label{flux}
\end{figure}

Additionally, enhanced Raman signals were observed between 90 and 150~cm$^{-1}$ in some \FMO crystals [red spectrum inside a dashed box in Figure~\ref{flux}]. These intensities were significantly weaker at different positions on the identical crystal, while other peaks remained unchanged [blue spectra in Figure~\ref{flux}]. Similar Raman peaks were also observed in a few \NZMO crystals, indicating that these peaks were not intrinsic. These peaks are likely from the residual flux, corresponding to three Te phonons (E, A$_{1}$, and E modes from left to right)~\cite{Yannopoulos2020}, i.e., polycrystalline Te powder could be generated from the decomposition of the transport agent TeCl$_{4}$ during the growth.

\section{Conclusions}
\label{summary}
We systematically examined the CVT method by growing two representative pyroelectric polar magnets, \FMO and \NZMO, which are also altermagnetic candidates. Through careful optimization of growth parameters, we identified that the transport agent gas density is key to growing large crystals. To grow large single crystals, the mid setup was found to be more suitable for \FMO, by suppressing convection kinetics while minimizing the harmful quartz reaction. On the other hand, the end setup was more suited for \NZMO to fully utilize the enhanced convection kinetics, where an increased transport gas density was beneficial for growing very big crystals. Based on this alternative control of kinetics, the smaller diameter of ampules for \FMO and larger ampules for \NZMO facilitated the growth of large single crystals, respectively. The high quality of the single crystals was confirmed by various experimental techniques. The manipulation of the magnetic properties upon nonmagnetic Zn doping was also shown in \NZMO. The deep understanding and effective control of the CVT growth conditions found in this study provide essential insights into quantum solid synthesis. This can advance the polar magnetic oxide studies for altermagnetic and multiferroic applications and the investigation of various other functional quantum materials in future studies.

\section{Acknowledgments}
This work was supported by the Institute for Basic Science (IBS-R011-Y3, IBS-R034-Y1, and IBS-R034-D1) and the Advanced Facility Center for Quantum Technology at Sungkyunkwan University. Part of this study was performed at facilities in the IBS Center for Correlated Electron Systems, Seoul National University. Access to the X-ray facilities at the Materials Characterisation Laboratory at the Rutherford Appleton Laboratory is gratefully acknowledged. The work at Rutgers University was supported by the DOE under Grant No. DOE: DE-FG02-07ER46382.

\bibliography{Reference}

\end{document}